\documentclass[amsmath,caption,amssymb,superscriptaddress,12pt,a4paper,hidelinks]{article}
\usepackage{graphicx}
\usepackage{color}
\usepackage{mathrsfs}
\usepackage{amsmath,amssymb}
\usepackage{txfonts}       
\usepackage[a4paper, total={6in,9in}]{geometry}
\usepackage[sort&compress,super]{natbib}
\setcitestyle{citesep={,}}
\usepackage{doi}
\usepackage{bm}
\usepackage{xcolor}
\usepackage{notes2bib}

\usepackage[normalem]{ulem}  
\usepackage{array}

\usepackage[font=footnotesize,labelfont=bf]{caption}

\usepackage{tabularx}

\title{Comprehensive view of microscopic interactions between DNA-coated colloids
}

\author{\normalsize{Fan Cui,$^{\dagger,1}$ Sophie Marbach,$^{\dagger,2,3}$ Jeana Aojie Zheng,$^{1}$  Miranda Holmes-Cerfon,$^{2}$} \\
\normalsize{and David J. Pine$^{\star,1,4}$}\\
\\
\small{$^{1}$Department of Physics, New York University, New York, NY, USA}\\
\small{$^{2}$Courant Institute of Mathematical Sciences, New York University, New York, NY, USA} \\
\small{$^{3}$CNRS, Sorbonne Universit\'{e}, Physicochimie des Electrolytes et Nanosyst\`{e}mes} \\
\small{Interfaciaux, F-75005 Paris, France}\\
\small{$^4$Department of Chemical \& Biomolecular Engineering, New York University,} \\
\small{New York, NY, USA}\\
}
\date{}

\begin{document}



\baselineskip24pt


\vspace{-2cm}

\maketitle
\newpage
\begin{abstract}
\linespread{1.6}
\normalsize{\bf

The self-assembly of DNA-coated colloids into highly-ordered structures offers great promise for advanced optical materials.
However, control of disorder, defects, melting, and crystal growth is hindered by the lack of a microscopic understanding of DNA-mediated colloidal interactions.
Here we use total internal reflection microscopy to measure \textit{in situ} the interaction potential between DNA-coated colloids with nanometer resolution and the macroscopic melting behavior.
The range and strength of the interaction are measured and linked to key material design parameters, including DNA sequence, polymer length, grafting density, and complementary fraction.
We present a first-principles model that quantitatively reproduces our experimental data without fitting parameters over a wide range of DNA ligand designs.
Our theory identifies a subtle competition between DNA binding and steric repulsion and accurately predicts adhesion and melting at a molecular level.
Combining experimental and theoretical results, our work provides a quantitative and predictive approach for guiding material design with DNA-nanotechnology and can be further extended to a diversity of colloidal and biological systems.
}
\end{abstract}


\newpage


DNA-coated colloids are our most versatile tool for the targeted self-assembly of colloidal materials~\cite{mirkin1996dna,alivisatos1996organization,park2008dna,nykypanchuk2008dna,macfarlane2011nanoparticle,he2020colloidal}, which have important applications in photonics\cite{colloidalphotonicreview1999,2001onchipassembly, color2015PNAS}, metamaterials\cite{XZhangMeta2014,metasurface2021}, and biomedical devices\cite{biosensing2019quantitative,drugdelivert2014dna}.
Their versatility stems from the programmability of their single-stranded DNA sticky ends, a sequence of nucleobases that enable specific attractive binding to particles coated with complementary ssDNA.
While programming the sticky ends determines which DNA ligands on different particles bind, control of collective processes, such as the binding-unbinding transition, commonly referred to as melting, and crystal growth and size, are hindered by our limited understanding of the microscopic mechanisms at play.
Recent experimental work, for example, has highlighted the importance of the areal density of the DNA ligands for controlling binding kinetics and optimizing crystallization\cite{chaikinkinetics,johncrockerswelling,WangbrothersJACS}.
Moreover, with the advent of patchy particles\cite{2012WangPatches,sacanna2017gongfusion} and colloids with complex non-spherical shapes\cite{2017ducrotNatMat,mirkin2017clathrate,he2020colloidal}, controlling the range of the interaction by adjusting the lengths of the ligands becomes increasingly important.
Predicting how all these parameters interact to control the binding and unbinding of DNA-coated particles represents a formidable challenge, which we address in this paper.
Directly probing microscopic DNA mediated interactions is an indisputable challenge~\cite{rogers2011direct,angioletti2019understanding,merminod2021avidity}.
This is due in part to the zoology of forces beyond DNA ligand binding occurring at these scales, including steric repulsion~\cite{milner1988compressing}, van der Waals\cite{parsegian2005van}, and electrostatics\cite{nykypanchuk2007dna,hueckel2020ionic}.
Early foundational work highlighted the crucial role of entropy during binding, which is related to ligand conformations\cite{dreyfus2009simple,rogers2011direct} and the competition between binding partners\cite{mognetti2012predicting,varilly2012general,angioletti2019understanding,angioletti2013communication,merminod2021avidity}.
However, limited investigation of different experimental designs makes it hard to pinpoint and control the interaction mechanism.
In particular, at high ligand coverage, which is used in most current designs and is needed for fast kinetics\cite{WangbrothersJACS,manoharan2016dnaccNatRevMat}, no direct comparison between experimental potential profiles and theoretical predictions exists.
In fact, standard modeling based on a discrete numerical account of ligands is intractable at high coverage due to the high number of ligands~\cite{rogers2011direct,merminod2021avidity}.
Finally, simultaneous measurement of microscopic interactions and macroscopic material properties remains uncharted.

In this paper we directly measure the interaction potential between high-density DNA-coated colloids using total internal reflection microscopy (TIRM).
We present first-principles modeling that completely reproduces experimental data over a wide range of single stranded DNA (ssDNA) ligand designs without any fitting parameters.
Experiments and theory highlight the crucial role of steric repulsion between brushes, balancing attractive binding forces over a short distance to form an extremely narrow potential well.
We demonstrate quantitative tuning of experimental systems to achieve high control over macroscopic material properties such as the melting temperature.
Finally, we unveil a microscopic view of binding, where surfaces start interacting with about a dozen bonds at the melting temperature, then strongly compress the polymer brush by up to 20~nm upon cooling.


\begin{figure}[h!]
    \center
    \includegraphics[width = 0.95\textwidth]{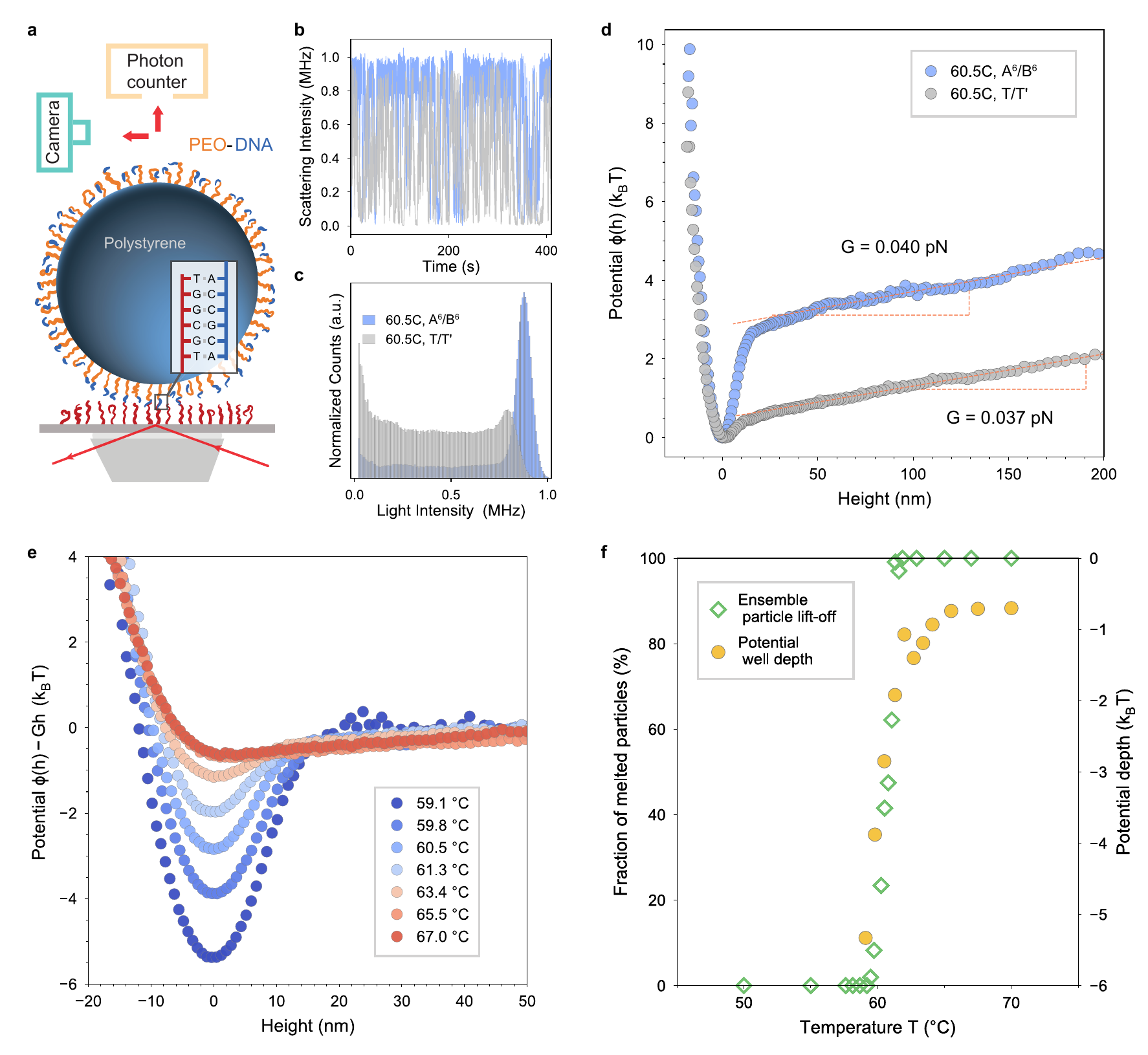}
    \caption{
        \textbf{Measurement of potential between DNA-coated surfaces using total internal reflection microscopy}. \textbf{a}, Schematic of the experimental setup.
        A DNA-coated PS particle is illuminated by an evanescent wave generated by a 633~nm laser.
        The scattered light is detected by a PMT photon counter and simultaneously with a camera to track particle positions.
        The glass substrate surface is coated with the complementary DNA strands, here with 6 sticky bases.
        Drawings are not to scale.
        \textbf{b}, Scattering intensity, \textbf{c}, corresponding statistical histogram and \textbf{d}, potential energy profile of a DNA-coated particle measured at $60.5^\circ$C.
        In \textbf{b-d}, particle/substrate pairs with DNA strands with 6 sticky bases, A$^6$/B$^6$, (respectively with non-interacting bases, T/T$^\prime$) are blue (resp. gray).
        \textbf{e}, Potential energy profiles, with gravity removed, of a DNA-coated particle (A$^6$/B$^6$) at different temperatures.
        \textbf{f}, Potential well depth vs.\ temperature (yellow circles) with superimposed melting curve, based on $\sim$200 particles (green diamonds) of DNA-coated particles (A$^6$/B$^6$).
        All particles measured have a diameter of $5~\mu$m and PEO linkers with molecular weight 34~kg/mol clicked to 20-base DNA strands.
        The glass is coated with 60-base DNA strands.
        Measurement solution consists of 140 mM PBS (pH=7.4) and 0.3\% F127.
        Refer to Table \ref{tab:DNAsquence} for full DNA sequences.}
    \label{figure1}
\end{figure}

\paragraph{Experiments.}
We probe the interaction potential between DNA-coated surfaces using a cus\-tom-built, highly sensitive total internal reflection microscope (Fig.~\ref{figure1}a).
Both surfaces of the poly\-styrene (PS) particle and the glass substrate are coated with ssDNA (Methods)~\cite{oh2015high, silaneazide}.
On the PS particle, the DNA strands are anchored onto the surface through a polyethelyne oxide (PEO) linker with variable molecular weight $M_w$.
The brush-mediated DNA functionalization results in a high-density DNA coating with densities ranging from 0.1 to 0.02~nm$^{-2}$ (Supplementary \S 2.2).
Interactions between the two surfaces are modulated both by varying $M_w$ of the PEO brush and the binding strength of the ssDNA sticky ends.

To measure the interaction potential using TIRM, a particle is illuminated with an exponentially damped evanescent wave.
As the particle travels vertically, the intensity of the light scattered by the particle, which decays exponentially with particle height\cite{prieve1990total,prieve1999measurement}, is measured with a photomultiplier (PMT) photon counter (Fig.~\ref{figure1}b).
The statistical distribution of the scattered intensities (Fig.~\ref{figure1}c) is therefore a measure of height distributions $p(h)$, which is related to the potential $\phi(h)$ by Boltzmann's equation:
\begin{equation}
    p(h) = \frac{1}{Z} \exp \left[-\frac{\phi(h)}{k_B T}\right] \;,
\label{eq:boltzmann}
\end{equation}
where $k_B$ is Boltzmann's constant, $T$ the temperature, and $Z$ a normalization factor.
When the number of observations is large (more than 350,000 in our experiments), $\phi(h)$ can be reliably inferred from light intensity distributions (Methods and Supplementary \S 1).

We employ the TIRM technique first to investigate colloid/substrate coatings with complementary (A$^6$/B$^6$, 6 sticky bases) and non-interacting (T/T$^\prime$) DNA strands (Fig.~\ref{figure1}a-c).
For a sample with ssDNA sticky ends, the scattered intensities (blue trace in Fig.\ \ref{figure1}) exhibit long intervals of high scattered intensity and relatively short intervals of low intensity, indicating that the particle spends most of its time bound to the surface and occasionally breaks bonds and diffuses away from the surface.
TIRM analysis of this signal, shown in Fig.\ \ref{figure1}d, shows that the sticky particle exhibits a sharp attractive potential well with a depth $\simeq 2.8~k_BT$ at small separation distances.
By contrast, a particle  with non-interacting strands spends much more time away from the surface and exhibits a very shallow well with a depth of less than $0.5~k_BT$.
The sharp attractive potential well for the particle with complementary strands can thus be attributed to DNA hybridization interactions.
By controlling the rate of photon detection, we confirm that the well width is broadened by shot noise from photon counting (Supplementary Fig. S15)\cite{cui2021tirmnoise}.
At larger separations beyond 50~nm, both potentials show a linear upward increase, consistent with the 0.037~pN gravitational force expected for our 5-$\mu$m diameter PS particles.

The temperature sensitivity of DNA hybridization is key for macroscopic assemblies.
Figure~\ref{figure1}e shows the potential profiles of a single particle with 6 sticky bases over an $8^\circ$C temperature range.
Here, the gravitational contribution is subtracted to emphasize surface interactions (Methods).
The attractive well becomes shallower as temperature is increased, indicating fewer hybridization bonds.
Figure~\ref{figure1}f shows that the potential well depth decreases from $5.5~k_BT$ to $0.8~k_BT$ between $59^\circ$C and $64^\circ$C.
As temperature increases above $64^\circ$C, the well depth plateaus around 0.8~$k_BT$. Interestingly, we still observe a non-zero attractive interaction potential at high temperatures when DNA hybridization should be negligible.
This potential well is also present in the non-interacting particle potential (Fig.~\ref{figure1}d).
This range and strength of the attraction is consistent with van der Waals interactions (Supplementary \S 3.5). 

We can relate the well depth, a microscopic single-particle property, to the melting of the DNA-coated particles, an important macroscopic material property. The melting of DNA-coated colloids aggregates is usually characterized by the fraction of unpaired particles (singlets) as a function of temperature~\cite{PNASmelting,WangbrothersDNAcolloids}. Here, we adopt a similar definition and plot the fraction of \textit{melted} particles as a function of temperature.
In the particle-substrate geometry, we take a particle as melted when it has lifted off from the surface, at least once 20~nm beyond the potential minimum during a 1-minute observation window (Supplementary \S 1.4). This method directly captures unbinding by measuring particle separation, in contrast to other work that infers melting by tracking lateral particle motion~\cite{xu2011subdiffusion}. We observe $\sim$200 DNA-coated particles using the camera on TIRM and plot the percentage of melted particles as a function of temperature in Fig.~\ref{figure1}f. The melting curve shows a sharp transition with $T_m \simeq 60.5^\circ$C, which coincides with the potential well depth of roughly $3~k_BT$. At potential depths above $1.5~k_BT$, the particles are completely melted. Figure \ref{figure1}f shows a clear correspondence between microscopic interaction energy and macroscopic ensemble melting.


\paragraph{Multiscale model.}
To understand how microscopic material design affects macroscopic melting, we build a predictive model.
A careful account of the polymer brush is central for quantitative description~\cite{mognetti2012predicting,varilly2012general,angioletti2013communication}.
While previous work relied on stochastic simulations, here the high densities render such approaches impractical and inaccurate~\cite{merminod2021avidity,rogers2011direct}.
Instead, we aim for a mean field description. A series of calibration experiments~\cite{youssef2020rapid} reveals PEO coatings that are thick, 12--40~$\mathrm{nm}$, and dense, 3--7~nm between grafts, (Supplementary \S 2), which validates the use of a model based on the Milner-Witten-Cates (MWC) brush model~\cite{milner1988compressing,milner1988theory,drobek2005compressing}.

\begin{figure}[t!]
    \center
    \includegraphics[width = 0.95\textwidth]{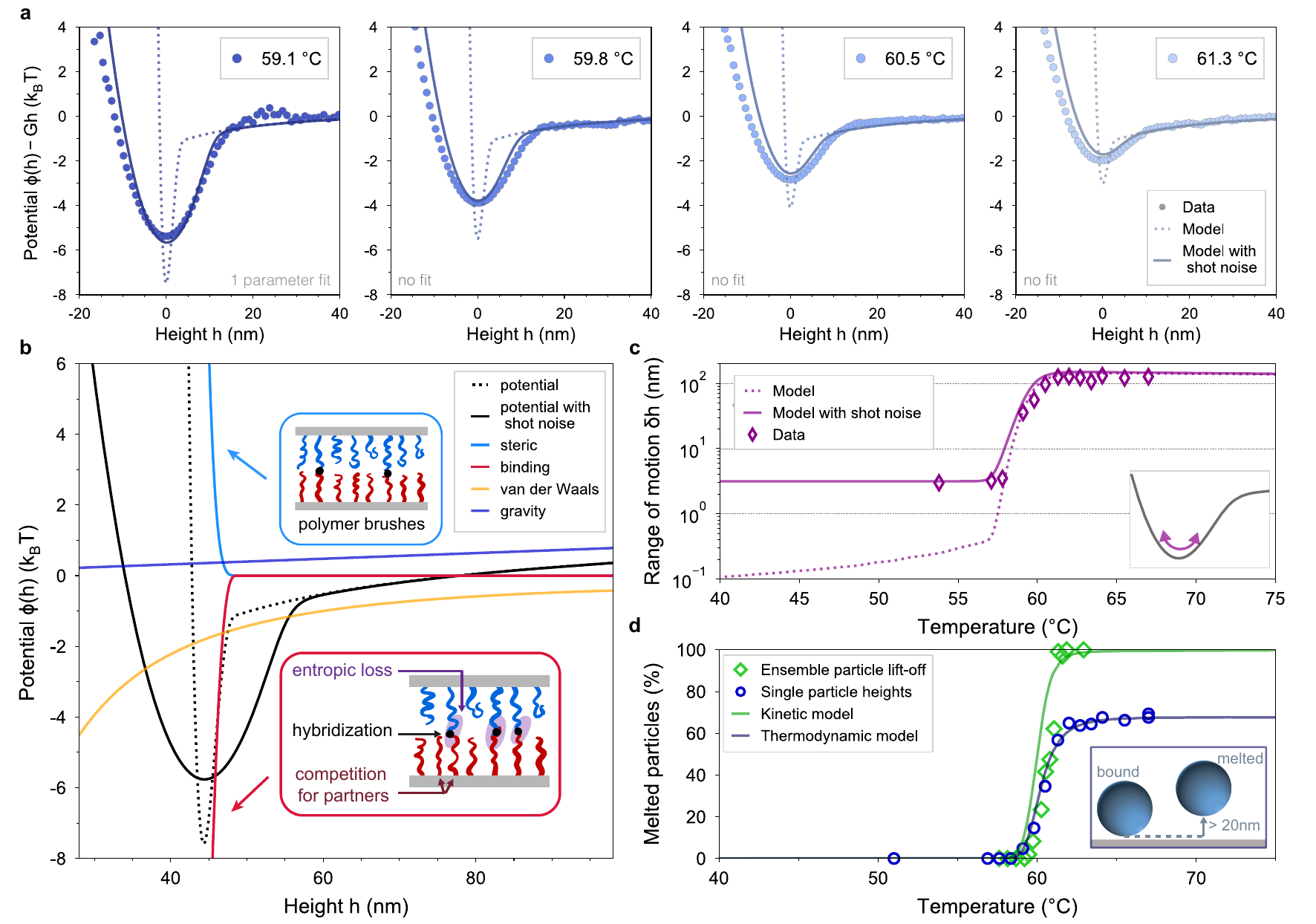}
    \caption{
        \textbf{Sticky polymer brush model reproduces experimental data}:
        \textbf{a}, Model and experimental potential profiles of the DNA-coated particle presented in Fig.~\ref{figure1}d with a measured coating density $\sigma = 0.021~\mathrm{nm}^{-2}$ of 34~k PEO with 6 sticky bases.
        The potential is shown at 4 temperatures around $T_m = 60.5^\circ$C.
        The density of strands on the glass ($\sigma_g = 0.011~\mathrm{nm}^{-2}$) is the only fitting parameter and is fitted only once at $59.1^\circ$C.
        No additional fitting is done for higher temperatures.
        Shot noise consistent with experimental measurements is applied to the model predictions.
        \textbf{b}, Contributions to the model potential profile at $59.1^{\circ}$C.
        (Insets) Schematics of components accounted for in the model.
        \textbf{c}, Range of motion $\delta h = \langle h^2 \rangle_{\phi}^{1/2}$ around the potential minimum, (inset) describing the typical displacement of the particle in the well (Supplementary \S 3.2.1).
        \textbf{d}, Experimental and model thermodynamic and kinetic melting curves. The experimental kinetic curve corresponds to the ensemble particle lift-off also reported in Fig.~\ref{figure1}e).
        To compute the kinetic melting curve we use the hindered diffusion of the free particle at height $h \simeq 0.3$~nm above the hydrodynamic floor, $D \simeq D_0 h/a$~\cite{bevan2000hindered} where $D_0 = 2.09 \times 10^{-13}$ m$^2$/s is the measured bare diffusion coefficient and $2 a = 5~\mathrm{\mu m}$ is the diameter of the particle.
        Both methods are insensitive to measurement noise (Supplementary \S 4.2).
        (Inset) The thermodynamic method measures the fraction of particles lifted at least 20~nm above their equilibrium position.
    }
    \label{figure2}
\end{figure}

We extend the MWC theory to our more complex brushes.
Our calculation of the steric repulsion $\phi_{\rm steric}(h)$ includes the effects of having a fraction $f$ of bound ends\cite{meng2003telechelic}, the heterogeneous composition of the tether (DNA--PEO), and the different DNA coating densities on the particles and the substrate~\cite{shim1990forces}.
The binding attraction $\phi_{\rm binding}(h)$ includes the free energy of hybridization of $\Delta G^{(0)}$ of the complementary DNA strands used~\cite{santalucia1998unified}.
Competition for binding partners and entropic contributions due to the loss of degrees of freedom upon binding are included consistently within our brush model~\cite{mognetti2012predicting,varilly2012general,angioletti2013communication}.
The fraction of bound ends $f$ is then determined by minimizing~$\phi_{\rm binding}(h) +  \phi_{\rm steric}(h)$ (Supplementary \S 3), in contrast with previous work that did not consider how bound ends modify steric repulsion~\cite{rogers2011direct,merminod2021avidity,mognetti2012predicting,varilly2012general,angioletti2013communication}.
Adding gravity and van der Waals attraction, the potential $\phi(h) =  \phi_{\rm grav} + \phi_{\rm binding} + \phi_{\rm steric}+  \phi_{\rm vdw}$, is shown by the dotted black line in Fig.~\ref{figure2}b. The experimental shot noise, which cannot be eliminated, can be quantitatively controlled to a known level\cite{cui2021tirmnoise}. Quantitative accounting for the Poisson-distributed photon counting effect yields the noise-broadened potential curve, shown by the solid black curve in Fig.~\ref{figure2}b (Supplementary \S 4).

Below the melting temperature, competition between binding attraction (red) and steric repulsion (blue) forms a narrow potential well, barely $2$~nm wide. Both contributions occur at the same separation: when brushes touch, binding (attraction) is favored as well as compression (repulsion).
The balance of these two forces is therefore subtle, and only observed around and below the melting temperature when the hybridization energy $\Delta G^{(0)}(T)$ is sufficiently favorable.
Importantly, our predictions of narrow potentials are well represented by new short range Lennard-Jones potentials~\cite{wang2020lennard} (and not by \textit{e.g.}\ Morse, Supplementary \S 5), validating their applicability for self-assembly simulations~\cite{angioletti2019understanding,das2021variational}.

\paragraph{Comparison of model and data.}
We use this model to compare the predicted potential profiles $\phi(h)$ to the experimentally measured ones.
Input parameters for the model, including temperature, brush length, and DNA density, are all taken directly from measured values (Table 1).
The hybridization energies of sticky ends are determined from tabulated values using the second nearest neighbor model of SantaLucia~\cite{santalucia1998unified}.
The only parameter that is not precisely known is the glass coating density $\sigma_g$, which is determined by fitting the potential to data obtained at a single temperature, $59.1^{\circ}$C, for a single particle type; the data and fit are shown in Fig.~\ref{figure2}a (first panel). We obtain $\sigma_g = 0.011~\mathrm{nm}^{-2}$, consistent with the expected range (Supplementary \S 2.2), and fix it at this value for all remaining predictions.

Figure \ref{figure2}a (remaining 3 panels) compares the measured potentials with model predictions at three different temperatures.
The predicted potentials are in remarkable agreement with experiments, including the temperature-dependence of the noise-adjusted widths and depths of the potential, as well as the overall shapes, particularly when one considers that there are no fitting parameters, apart from the calibration experiment.
Interestingly, while the measured potential width is about $10~\mathrm{nm}$ (defined as the full width at $\simeq 1~k_B T$ above minimum) at 59.1 $^\circ$C, our model indicates that it is broadened by shot-noise from a potential $\sim$ $2~\mathrm{nm}$ wide.
Such narrow widths are consistent with separate investigations of effects of photon counting shot noise on TIRM measurements\cite{cui2021tirmnoise}.

Particle unbinding can be directly probed from potential measurements by plotting the root mean square height fluctuations $\delta h= \sqrt{\langle h^2 \rangle_\phi}$, around the potential minimum (see Figure \ref{figure2}c.)
The rms height $\delta h$ quantifies the range of motion of the particle and can be calculated from both our TIRM measurements and our model.
For temperatures below $57^\circ$C, the particle is tightly confined within the potential well with a predicted range of motion $\delta h_{\rm} \lesssim 0.4$~nm; the TIRM measurements saturate at $\delta h_{\rm} = 3$~nm, a value set by the photon counting shot noise.
Clear evidence of particle unbinding is seen at $58^\circ$C and above, where $\delta h$ increases dramatically over a temperature window a few degrees wide.
At temperatures above $62^\circ$C, $\delta h$ reaches a plateau set by  the gravitational height of the particle, which is 124~nm (Supplementary \S 3.2.1).

Macroscopic quantities such as the melting temperature $T_m$ can also be modeled. From a modeling perspective, the ``lifting-off'' criteria for the ``melted'' particles discussed earlier is essentially a \textit{kinetic} definition and it corresponds to a mean first passage time problem for the particles to lift-off beyond the potential well~\cite{gardiner1985handbook} (Supplementary \S 3.2). The \textit{kinetic} melting curve depends on the vertical diffusion coefficient of the particle. Here we approximate the diffusion coefficient by a representative value corresponding to hindered hydrodynamic diffusion near a wall~\cite{brenner1961slow} and find excellent agreement with the experimental measurement (see Fig.~\ref{figure2}d).

\begin{figure}[h!]
    \center
    \includegraphics[width = 0.65\textwidth]{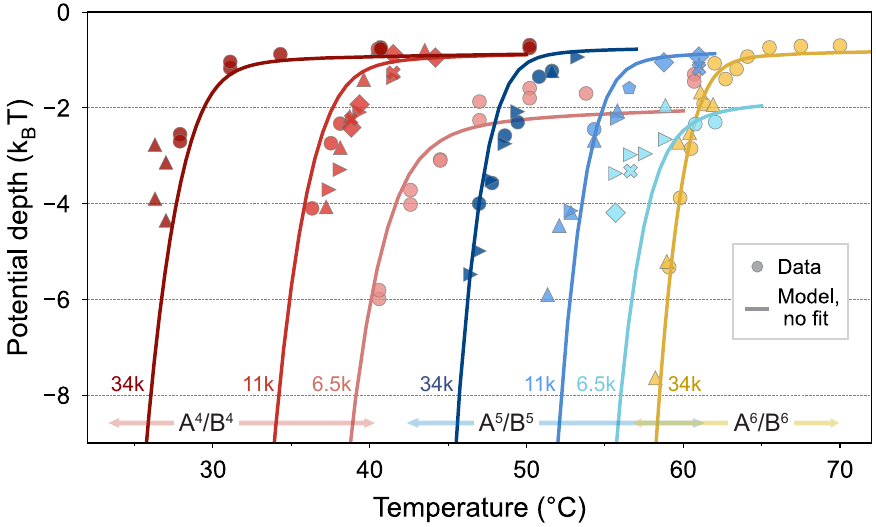}
    \caption{
        \textbf{Control of melting properties with hybridizing pair number and chain length}:
        Potential well depths measured for different experimental systems: PEO brushes with $M_w$ of 6.5~k, 11~k and 34~k, and colloid/glass surfaces attached with 4 (-GCAG/-CTGC), 5 (-GACGC/-GCGTC) or 6 (-TGCGGT/-ACCGCA) DNA complimentary bases.
        The different symbols correspond to different particles.
        The model predictions (with no additional fitting) with photon shot noise are overlaid.
        The particles with 34~k and 6 sticky bases correspond to that of Fig.~\ref{figure1}e and f and Fig.~\ref{figure2}. At high temperatures, all particles experience a constant $\sim$1 -- 2 $k_B T$ well depth from van der Waals interaction.
        }
    \label{figure3}
\end{figure}

To provide another perspective on melting, we introduce a single-particle, \textit{thermodynamic} melting definition, where the fraction of melted particles corresponds to the fraction of time the particle remains unbound at heights $h$ beyond the attractive binding well ($h \gtrsim20~\mathrm{nm}$):
\begin{equation}
    p_{\rm unbound} = 1 - \frac{1}{Z} \int_0^{h_c} \exp \left[ -\frac{\phi(h)}{k_B T} \right] dh.
\label{eq:punbound}
\end{equation}
This definition, though on a single-particle level, captures the same thermodynamic melting picture drawn from a large ensemble of particles that rely on statistics to determine the fraction of bound particles. The single-particle melting curve is obtained from the raw scattering data (e.g. Fig 1b) that traces the particle position during the one measurement and is shown in Fig.~\ref{figure3}c (blue circles). The model prediction (line), shows excellent agreement with experiments. Note that $p_{\rm unbound}$ does not reach $100\%$ at high temperatures due to gravity.
Remarkably, the kinetic and thermodynamic melting curves transition at the same temperature $T_m = 60.5^{\circ}$C. 
This sheds light on the melting dynamics: DNA pairs unbind and a particle lifts-off with an escape rate of the order of 1~min (Supplementary \S 3.2.2).


\paragraph{Control of macroscopic melting.}

Our model indicates that the key to control the melting point is to adjust the balance between hybridization and steric repulsion.
Hybridization is determined by the type and number of the sticky bases, while steric repulsion is determined by the length and density of the underlying PEO polymer layer.
To test our model, we experimentally measured the potential well depth of colloids with different number of complementary DNA sticky bases (4, 5, or 6) and PEO molecular weights ($M_w = 6.5$, $11$, and $34$ kg/mol). For each configuration, multiple particles are measured to ensure the reproducibility and homogeneity of our measurements. The experimental as well as the calculated well depth results are shown in Fig.~\ref{figure3}. 

First, we find that our theoretical predictions are in close agreement with the well depths measured experimentally with TIRM, for a wide variety of colloid coatings with no fitting parameters.
The robust agreement, spanning a broad range of temperatures, indicates that our model is able to successfully capture the detailed molecular balance that determines macroscopic melting.

Second, we find that changing the binding attraction by reducing the number of sticky bases from 6 (yellows) to 5 (blues) to 4 (reds) consistently depresses the melting point by 10 to 30$^{\circ}$C for each PEO length.

Third, we observe that for a given number of sticky bases, decreasing the chain length from $M_w = 34$ kg/mol to $6.5$ kg/mol consistently increases the melting point by up to 15$^{\circ}$C.
Separate model inquiries (Supplementary \S 6) show that this originates equally from two contributions.
Decreasing chain length decreases the number of polymer strands whose degrees of freedom are reduced by compression of the brush.
This reduces the entropic penalty of compressing the brush, which decreases steric repulsion and promotes binding.
Moreover, experimentally we find that shorter chain lengths are associated with higher areal densities.
Our model shows that increased areal density increases binding attraction more than it increases steric repulsion, which is a useful and interesting insight (Supplementary \S 6).

The faithful reproduction of experimental data supports its use as a predictive tool for the rapid exploration of different material designs\bibnote{To enhance accessibility, our model is available through GitHub: https://github.com/smarbach/DNACoatedColloidsInteractions}.
In fact, we also find close agreement of melting temperature predictions with previously reported experimental data of different material systems by other researchers~\cite{rogers2011direct,xu2011subdiffusion}, including at lower coating densities (Supplementary \S 7). Additionally, our mechanistic knowledge suggests new design avenues. For example, two different DNA sticky sequences could  melt at the same temperature (as is nearly obtained here for $5$ bases, $M_w = 6.5$ kg/mol and $6$ bases, $M_w = 34$ kg/mol) or even at inverted temperatures, by tuning the length of the polymer chain, thus permitting a change in the range of interaction without necessarily changing its strength.


\paragraph{Melting and binding.}

Our TIRM measurements uncover a detailed microscopic picture of DNA-mediated melting and binding below the melting temperature.
By monitoring the separation between binding partners at the nanometer scale, TIRM provides detailed information about binding beyond the 2-degree melting transition window probed by macroscopic melting curves as it also measures how the polymer brush is compressed when the temperature is lowered far below the melting temperature.

\begin{figure}[h!]
    \center
    \includegraphics[width = 0.95\textwidth]{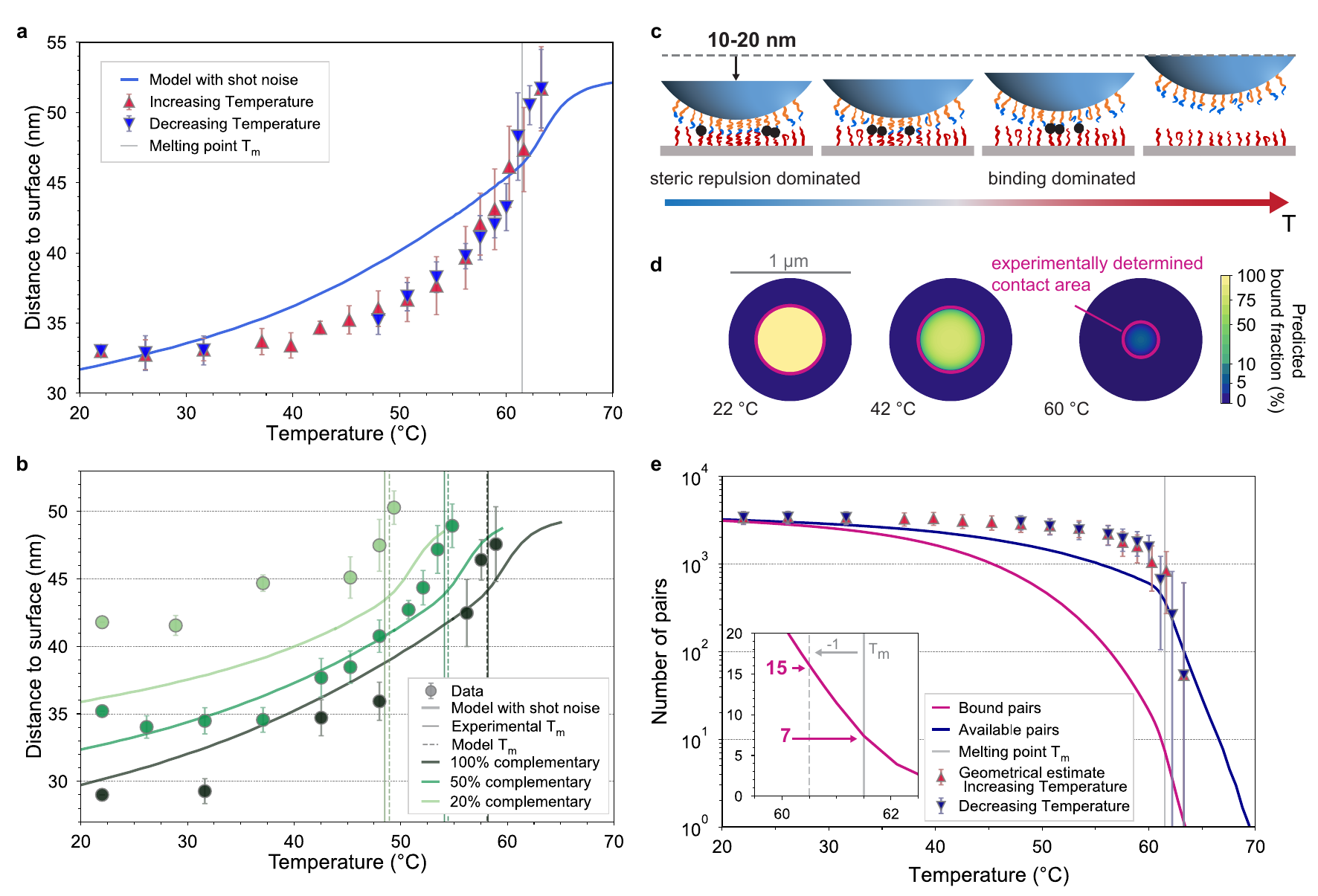}
    \caption{
        \textbf{Microscopic insight into the melting transition}:
        \textbf{a}, Height $h$ of potential minimum for a 34~k PEO + A$^6$/B$^6$ particle, measured with TIRM for increasing and decreasing temperatures and model predictions.
        For this system $T_m = 61.5^{\circ}$C.
        \textbf{b}, Height of particles with varying complementary coverage fraction $x$, with 34~k PEO + A$^6$/B$^6$ ($x\%$) + T/T$^\prime$ ($100-x\%$), as measured with TIRM for decreasing temperatures.
        $T_m = 59.2^\circ$C for the 100\% complementary.
        In the model a particle density $\sigma_p = 0.012~$nm$^{-2}$ (within the measured density variability) was used to yield the same $T_m$ as measured for 100\%.
        No additional fitting is done for the 20 or 50\%.
        \textbf{c}, Sketch of microscopic melting from low to high temperatures.
        Bonds (which fluctuate in time) are highlighted with black dots.
        \textbf{d}, Contact area with radius $r \simeq 2 a (h_{\rm max} - h)$ where $h$ is the separation distance from \textbf{a} and $h_{\rm max} \simeq 52~\mathrm{nm}$ the maximal interaction height. Here $2a = 5~\mathrm{\mu m}$. Overlapped model predictions show the proportion of bound pairs over the particle's bottom surface.
        \textbf{e} Number of pairs within range (available) for binding calculated from experimental data of \textbf{a} as $\pi r^2 \sigma_g$ with $r$ from \textbf{d}. Error bars correspond to $6\%$ uncertainty on $h_{\rm max}$.  Model prediction for the average number of available and bound pairs and (inset) zoom around $T_m$ (Supplementary \S 3.7).
        Model parameters for \textbf{d} and \textbf{e} correspond to those of \textbf{a}.}
    \label{figure4}
\end{figure}

Figure \ref{figure4}a shows TIRM measurements of the height $h$ of the potential minimum from $6^\circ$C above to $40^\circ$C below $T_m$ for a 34~k PEO + A$^6$/B$^6$ particle-substrate interaction.
The measurements reveal a two-stage binding process.
Within a few degrees above and below $T_m$, $h$ decreases rapidly as the particle binds and the polymer brushes are compressed 10~nm from about 52 to 42~nm over a $4^\circ$C temperature window.
Within this window, the particle dynamically binds and unbinds, spending an increasing fraction of time bound towards the lower end of this temperature range, until the potential depth reaches about $8~k_BT$.
Below this range, in the second stage of binding, the particle remains bound at all times and its vertical excursions are increasingly confined to a narrow sub-nanometer range around its mean position, as seen in Fig.~\ref{figure2}c.
However, in this second stage, the polymer brush continues to be compressed by nearly another 10~nm as the temperature is lowered until the height reaches a plateau value of about 33~nm at $T \simeq 40^\circ$C.
This corresponds to a coefficient of linear thermal expansion on the order of $400 \times 10^{-6}~\mathrm{K}^{-1}$, 10--100 times greater than metals and several times larger than most rubbers. No hysteresis is apparent as the temperature is lowered and raised, suggesting thermal equilibrium.

Model predictions show a similar quantitative trend with a 20-nm compression of the brush, but the transition is more gradual with less of a distinction between the two stages of binding and compression.
We attribute the discrepancy to the presence of an adsorbed short copolymer (Pluronic F127, Methods) within the brush, which makes the brush softer towards its edge (Supplementary \S 2.3).

We further investigate control of melting and binding by altering the relative strengths of attraction and repulsion.
To do so, we vary the fraction of polymer strands that have DNA sticky ends from $100\%$ to $50\%$ and $20\%$, which decreases binding attraction without affecting steric repulsion.
As shown in Fig.~\ref{figure4}b, all particles exhibit a two-stage height change with temperature, as we observed previously, including a sharp compression near melting followed by a more gradual compression toward an ultimate plateau at low temperatures.
Our model correctly predicts the melting temperature in all cases, but overestimates the compression of the brush for the case of 20\% DNA sticky end fraction.
This may be due to a breakdown of the mean field approximation as the fluctuations in the areal density of DNA sticky ends are significant in this case, meaning that some sticky ends may not be able to find a binding partner at low temperatures.

The observed compression of the polymer brush provides useful insights into the melting process.
Far below the melting temperature where the particle height is about 33~nm (Fig.~\ref{figure4}a), the brushes from the two surfaces are in contact over an area with a diameter of about 500~nm (Fig.~\ref{figure4}d).
Of the 3000 sticky-end pairs present in this area, all are hybridized at low temperature (Fig.~\ref{figure4}e), thus allowing new bonds only at the periphery.
As the temperature is increased, the area and number of sticky ends available for binding decreases; the fraction of available bonds that actually bind also decreases, as shown in Fig.~\ref{figure4}e.
At $T_m$, our model predicts 7 bound pairs out of about 300 pairs that are available for binding, which stabilizes the interaction.


\paragraph{Outlook.}

As particle designs and directed colloidal self-assembly targets become more ambitious, the need to understand and control the interaction range, strength, and cooperativity of DNA brushes tethered to particle surfaces becomes a pressing problem.
Here we provide a comprehensive understanding of DNA-mediated particle binding, which is directly probed by TIRM experiments and analytically described by detailed account of microscopic forces, wrapped in a mean-field, rapidly-computed polymer model. The results allow one to quantitatively predict how ligand length, density, and binding strength interact cooperatively to control melting and binding behaviors.
The cooperativity of these interactions make them highly nonlinear, as illustrated by how increasing areal density of ligands increases both the strength of DNA binding and steric repulsion, but with binding attraction winning due to nonlinearities.
The quantitative accuracy of our model means it can be used as a design tool to engineer colloidal interactions at the nanoscale~\cite{Note1}.
The physical picture also provides insights about the dynamics of binding, which are important for annealing and controlling diffusion of DNA-bound colloids.
Thus, the combined experimental and theoretical approach can serve as a platform for understanding and controlling the dynamics of DNA-mediated binding, an important frontier for the self-assembly of DNA-coated colloids\cite{lee2018modeling,xu2011subdiffusion,licata2007colloids}.
For example, exotic designs, with multiple DNA sequences and chain lengths on a single particle to enable binding with different strengths to other particles\cite{crocker2017distorteddiamond}, are a natural, albeit challenging, sequel.
Finally, our experimental and theoretical assay could serve as a stepping stone to understand dynamics of other multivalent ligand-receptor systems, such as virus motion and stalling on mucus through adhesive proteins\cite{sakai2017influenza,muller2019mobility}.

\section*{Methods}

All chemicals are purchased from Sigma-Aldrich unless otherwise specified and used as received.

\subsection*{Preparation of DNA-coated polystyrene colloids}

We synthesize DNA-coated polystyrene (PS) spheres using a previously reported polymer brush approach with minor modifications\cite{oh2015high}.
Polystyrene-b-poly(ethylene oxide) copolymer (PS-b-PEO) are first functionalized with azide at the end of the PEO chain (PS-b-PEO-N$_3$)\cite{Azide_hiki2007facile}.
Polymers with three different molecular weights ($M_w$) are studied:

\begin{enumerate}
    \item PS$_{3.8k}$PEO$_{6.5k}$, $M_w = 11,124$ g/mol, PS = 3800 g/mol, and PEO = 6,500 g/mol;
    \item PS$_{3.2k}$PEO$_{11k}$, $M_w = 15,336$ g/mol, PS = 3200 g/mol, and PEO = 11,000 g/mol;
    \item PS$_{3.8k}$PEO$_{34k}$, $M_w = 39,690$ g/mol, PS = 3800 g/mol, and PEO = 34,000 g/mol.
\end{enumerate}

PS-b-PEO-N$_3$ are then attached to PS particles using the swelling/deswelling method\cite{oh2015high}.
Typically, $250~\mu$L aqueous solution containing 0.005 g/ml particles and 0.5~$\mu$M PS-b-PEO-N$_3$ is mixed with 160 $\mu$L tetrahydrofuran (THF) at room temperature.
The mixture is placed on a horizontal shaker (1000 rpm) for 1.5 hours to fully $swell$ the PS particles and absorb the PS block of the PS-b-PEO-N$_3$ molecules.
Then THF is slowly removed from the solution via evaporation, leaving the hydrophobic PS blocks inserted into the particles and the hydrophilic PEO chains extending out into the solution. The particles are washed with de-ionized water three times to remove excess polymers.

Single stranded DNA (ssDNA, 20 bases, purchased from Integrated DNA Technologies with $5^\prime$ dibenzocyclooctyne (DBCO) end modification, is clicked onto PEO tips through strain promoted alkyne-azide cycloaddition (SPAAC)~\cite{oh2015high}. Briefly, PS particles (0.0025 g/ml) previously coated with PS-b-PEO-N$_3$ polymer brush are dispersed in 400~$\mu$L 500mM PBS buffer, pH=7.4. Then 10~$\mu$L of DBCO-DNA (0.1~mM) are added into the suspension. The mixture is left to react for 48 hours on a horizontal shaker (1000 rpm).
The final product is washed in DI water three times and stored in 140 mM phosphate-buffered saline solution (PBS, pH = 7.4) at $4^\circ$C. The DNA coverage density is measured using flow cytometry (Supplementary \S 2.2). During TIRM measurements, the colloids are further diluted in 140 mM PBS buffer with 0.3\% F127 added as surfactants.

\subsection* {DNA coating on glass substrate}

Polished glass slides ($25~\mathrm{mm} \times 75~\mathrm{mm}$, purchased from Delta Technologies) with indium tin oxide (ITO) coated on one side, are first cleaned by sonication (acetone and isopropanol, IPA) and oxygen plasma.
The cleaned glass slides are then immersed in a toluene solution containing 11-azidoundecyltrimethoxysilane (N$_3$-DTMOS, 2\% v/v) and 0.03 M N,N-diisopropylethylamine for 48 hours.
The silanization of the glass surface is carried out in a N$_2$ filled glovebox to avoid moisture induced uneven polymerization and multi-layer formation\cite{silaneazide}.
The substrates are then rinsed with toluene, acetone and finally IPA before annealed at 120 $^\circ$C for 30 min in the glovebox.

We then define the DNA coverage area on an azide-functionalized glass substrate (the glass side) using a silicon reaction cell ($22~\mathrm{mm} \times 22~\mathrm{mm} \times 0.6~\mathrm{mm}$).
2.5 $\mu$M DBCO modified ssDNA (60 bases) disolved in 500 mM PBS (pH = 7.4) solution are injected into the cell to react with the N$_3$-DTMOS on the glass surface.
The reaction is carried out with gentle horizontal shaking (200 rpm) for 48 hours.
Finally the substrate is rinsed in DI water for 10 min to remove the excess DNA before drying with a stream of N$_2$.

\subsection* {TIRM design and data acquisition}

We use a total internal reflection microscope (TIRM) to measure the potential energy of a DNA-coated PS sphere near a glass surface.
The microscope is designed and custom built in our lab.
A detailed optical train is included in Supplementary (Figure~S1).
Light from a linearly polarized 632.8-nm laser (HeNe, 30 mW, Lumentum) is directed toward the glass-water interface at $70^\circ$ incident angle to generate an evanescent field.
The laser is coupled to the measuring cell using an asymmeric dove-shaped glass prism ($63^\circ/70^\circ$, H-K9L, Tower Optical Corporation), with the $70^\circ$ side surface facing the incident beam and the top surface in contact with glass slide though immersion oil.

Scattered light is collected using a $50\times$ long working distance objective (Mitutoyo Apo Plan 50X, 0.55NA, 13mm WD). A plate beamsplitter (50:50, Thorlabs) is put in the collection path to split the collected light into two directions, with one going into a photomultiplier tube (PMT) photon counter (Hamamatsu, H7421-40) and the other forming images on a CMOS camera (AMScope, MU1803).
We use a National Instruments counter (USB-6341) to output the received photon frequencies. For the frequency counting, we use the ``target photon number'' method in which the counter counts up to a certain number of photons to calculate a frequency. The target photon number is typically set to be 1000.

\subsection*{Potential profiles from light intensity scattering with TIRM}

When a particle is illuminated by an evanescent field, its scatting intensity $I$ at certain height $h$ follows the exponential relationship:
\begin{equation}
    I(h) = I_0 \exp (-\alpha h),
\label{eq:evanescent}
\end{equation}
where $I_0$ is the scattering intensity when $h$ = 0 and $\alpha ^{-1} $ is the penetration depth of the evanescent field. In this work, all measurements are done with $\alpha^{-1} = 99$ nm.

During a TIRM measurement, a colloidal particle is settled close to the glass substrate and scatters light as it moves vertically via Brownian motion.
At equilibrium, the probability that the particle is at a height $h$ above the substrate is given by the Boltzmann distribution, Eq.~(\ref{eq:boltzmann}).
We typically set the lowest measured potential energy to be zero and the corresponding height to be $h_m$.
Then the potential energy can be written as $\phi(h)/k_BT = \ln [p(h)/p(h_m)]$. We infer $p(h)$ from the histogram statistic of the time-dependent scattering.
When the number of observation is large, which is typically more than 350,000 in our experiment, $p(h)/p(h_m) \simeq n(h)/n(h_m)$ where $n(h)$ is the number of times the particle stays in the vicinity of height $h$.
The probability of the particle being at a height $h$ is equal to the probability of measuring the $h$-corresponding intensity $I$: $p(I) |dI| = p(h)|dh|$, where $p(I)$ is the probability density of the light intensities observed. With Eq.~\ref{eq:evanescent}, $p(h) = p(I)\alpha I$.

Hence, the potential energy can be written as:
\begin{equation}
    \frac{\phi(h)-\phi(h_m)}{k_B T}= \ln \left[\frac{n(I_m)I_m}{n(I)I}\right].
\label{eq:potential}
\end{equation}
where $I$ is the scattering intensity when the particle is at height $h$, $n(I)$ is the number of observations of intensity in the range from $I$ to $I$ + $\delta I$. We refer the lowest potential energy $\phi(h_m)$ to be 0 and the corresponding height and scattering intensity to be $h_m$ and $I_m$ respectively.

\subsection*{Gravity removal and well depth determination}

To remove the gravitational contribution $Gh$ from the potential, we first acquire the gravity through linear regression of the potential curve in the large-separation, linearly increasing region (usually 80-200 nm above potential minimum). The fitted slope $\Delta \phi/ \Delta h$ = $G$. For potential curves originating from the same particle, we fit gravity and obtain a value of $G$ for each potential profile. Then take the average value of $G$ to remove gravity from all potential curves for this specific particle.

Potential well depths are calculated from potentials with gravity removed. We apply parabola fitting around the potential minimum within the potential well and compute the fitted minimum as the potential well depth. Parabola fitting range is usually done up to $h_m \pm 4 nm$.

\subsection*{Multiscale model}

The free energy of the DNA-coated colloid $\phi(h)$ is a sum of bulk (gravity) and surface contributions $\phi(h) = \phi_{\rm grav}(h) + \phi_{\rm surf}(h)$.
We account for surface contributions with a Dejarguin approximation~\cite{derjaguin1934untersuchungen} allowing us to relate $\phi_{\rm surf}(h)$ and the surface interaction energy $\varphi(h)$ of two flat walls separated by a distance $h$
\begin{equation}
    \phi_{\rm surf}(h) = 2 \pi a \int_{h}^{\infty} \varphi(h')\, dh'\;,
\label{eq:dejarguin}
\end{equation}
where $a$ is the radius of the colloid under consideration.
Framed as such, the model is applicable to other geometries, including colloid-colloid interactions where Eq.~\eqref{eq:dejarguin} is simply divided by a factor 2.
The surface interaction energy includes binding attraction, steric repulsion and non-specific interactions attributed to van der Waals forces $\varphi(h) =  \varphi_{\rm binding}(h) + \varphi_{\rm steric}(h) +  \varphi_{\rm vdw}(h)$.

To evaluate $\varphi_{\rm binding}(h)$, we start by calculating the free energy of hybridization of the free chains $\Delta G^{(0)}$, using standard methods~\cite{santalucia1998unified}. Entropic contributions strongly depend on the detailed characteristics of the polymer coating, and modify the free energy of binding as
\begin{equation}
    \Delta G^{\rm eff}(h) = \Delta G^{(0)} - k_B T \log \left( \frac{\sigma \int_0^h C^{\rm top}(z) C^{\rm bottom}(z) dz}{\rho_0}\right)
    \label{eq:DeltaGeff}
\end{equation}
where $C^{\rm top/bot}(z)$ are respectively the top and bottom concentration of sticky ends relative to the surface distance $z$ and $\rho_0$ is the elementary concentration. Here the lost degrees of freedom are essentially vertical as the brushes are tightly packed on the surface and well described by a brush (\textit{e.g.}\ vertical) model. We use a modified version of the Milner-Witten-Cates (MWC) theory to express $C^{\rm top/bot}(z)$ (Supplementary \S 3.3). It contains the detailed measured experimental parameters, such as the brush density $\sigma$, its molecular weight $M_w$, its persistence length and its measured equilibrium length. We finally write the free energy of binding as
\begin{equation}
    \varphi_{\rm binding}(h,f) = \sigma  f \Delta G^{\rm eff}(h) + \sigma \left[ f \log f + 2 (1-f) \log (1-f)  + f \right]
\label{eq:binding}
\end{equation}
where $f$ is the fraction of bound sticky ends and the added terms compared to $ \Delta G^{\rm eff}(h)$ account for competition between binding partners~\cite{angioletti2013communication}. In contrast with earlier theory works where the fraction of bound sticky $f$ is determined self-consistently (for example as the minimizer of $\varphi_{\rm binding}(h) $)~\cite{mognetti2012predicting,varilly2012general,angioletti2013communication}, here we must take into account the fact that bound polymers modify steric repulsion.

Steric repulsion between opposing polymer brushes is accounted for through a repulsive potential $\varphi_{\rm steric}(h,f)$. Standard theories for steric brush repulsion such as MWC~\cite{milner1988compressing,milner1988theory,drobek2005compressing} have to be carefully extended to acknowledge for the complexity of our brush. We account for the fraction of bound brushes $f$~\cite{meng2003telechelic}, heterogeneous brush composition (DNA strands clicked onto PEO) and coating asymmetries (the particle's brush and coating density is slightly different from the glass)~\cite{shim1990forces}. $\varphi_{\rm steric}(h,f)$ therefore includes steric contributions for hybrid, asymmetric brushes with bound fraction $f$, and thus includes the details of the glass (g) and the particle (p) coatings through their density $\sigma_{g/p}$, molecular weights $M_{w}^{\rm PEO}$, $M_{w,g/p}^{\rm DNA}$ and persistence length of the PEO and DNA parts. It also allows us to specify the concentration of sticky ends $C^{\rm top/bot}(z)$. Lengthy expressions and detailed methods are reported in Supplementary \S 3.2-3.4. Note that also Eqns.~\ref{eq:binding} is extended to account for asymmetric coatings, as reported in Supplementary \S 3.2-3.4. The fraction of bound ends $f$ is then obtained by minimizing $\varphi_{\rm binding}(h,f) +  \varphi_{\rm steric}(h,f)$ with respect to $f$.

\newpage

\section*{Acknowledgements}
We gratefully acknowledge Dr. Kaiyuan Yao for his help in the designing and building of our microscope.
The authors are also indebted to Alexandre Morin for his guidance on zeta potential measurements to evaluate the thickness of polymer brushes, and to Stefano Angioletti-Uberti for numerous discussions on the binding part of the energy of interaction. The authors would further like to acknowledge fruitful discussions with Guillaume Dubach, Jaeup Kim, Madhavi Krishnan, Bartolo M. Mognetti.
S.M. acknowledges funding from the MolecularControl project, European Union’s Horizon 2020 research and innovation programme under the Marie Skłodowska-Curie grant award number 839225. S.M., M.H.-C., J.A.Z., and F.C. were supported in part by the MRSEC Program of the National Science Foundation under Award Number DMR-1420073.
The research of F.C. and D.J.P. was primarily supported by the US Department of Energy DE-SC0007991 for the initiation and design of the TIRM experiments.
Additional support for F.C. and D.J.P. was provided by the US Army Research Office under award number W911NF-17-1-0328.
M.H.-C. was partially supported by the US Department of Energy under Award No. DE-SC0012296, and acknowledges support from the Alfred P. Sloan Foundation.

\section*{Author information}

These authors contributed equally: F.C. and S.M.

\section*{Author Contributions}

F.C. synthesized the materials, designed and built the TIRM, performed the experiments and developed data analyses of the measurements.
S.M. developed the model, calibrated the parameters and analyzed the generated data.
F.C. and J.A.Z. measured the DNA/brush coating density of the colloids and substrates.
J.A.Z. carried out the electrophoresis experiments.
D.J.P. and M.H.C. supervised the project.
F.C., S.M., and D.J.P. wrote the paper.
All authors discussed the results and commented on the manuscript.

\section*{Competing Interests}

The authors declare no competing interests.

\section*{Additional Information}
Correspondence and requests for materials should be addressed to D. J. P. (pine@nyu.edu).
\newpage

\bibliography{DNA}

\newpage
\section*{Tables}

\begin{table}[h!]
\footnotesize
\centering
\begin{tabular}{p{0.22\textwidth}>{\raggedright\arraybackslash}p{0.32\textwidth}}

\textbf{Strands} & \textbf{Sequence} \\
\hline
\hline

\multicolumn{2}{l}{\textit{DNA on colloids}} \\
\hline
Complementary A$^6$ &   $5^\prime$-/DBCO/-T$_{14}$-\textbf{\emph{ACCGCA}}-$3^\prime$\\
Complementary A$^5$ &  $5^\prime$-/DBCO/-T$_{15}$-\textbf{\emph{GACGC}}-$3^\prime$   \\
Complementary A$^4$ &  $5^\prime$-/DBCO/-T$_{16}$-\textbf{\emph{GCAG}}-$3^\prime$   \\
Non-interacting T &  $5^\prime$-/DBCO/-T$_{20}$-$3^\prime$   \\
\hline
\multicolumn{2}{l}{\textit{DNA on glass surfaces}} \\
\hline
Complementary B$^6$ &   $5^\prime$-/DBCO/-T$_{54}$-\textbf{\emph{TGCGGT}}-$3^\prime$\\
Complementary B$^5$ &  $5^\prime$-/DBCO/-T$_{55}$-\textbf{\emph{GCGTC}}-$3^\prime$   \\
Complementary B$^4$ &  $5^\prime$-/DBCO/-T$_{56}$-\textbf{\emph{CTGC}}-$3^\prime$   \\
Non-interacting T$^\prime$ &  $5^\prime$-/DBCO/-T$_{60}$-$3^\prime$   \\
\hline
\end{tabular}
\caption{\label{tab:DNAsquence} Sequence of the DNA strands used in this work}
\end{table}

\begin{table}[h!]
\footnotesize
\begin{tabular}{p{0.3\textwidth}>{\centering}p{0.3\textwidth}>{\raggedleft\arraybackslash}p{0.35\textwidth}}
\textbf{Parameter} & \textbf{Range of values} & \textbf{Calibration method}\\
\hline
\hline
\multicolumn{3}{l}{\textit{Measurement cell}} \\
\hline
Temperature & $T = 22 - 70~^{\circ}C$ & temperature control \\
Buffer salt concentration & $140~\mathrm{mM}$ & known \\
Solution density & $\simeq 1~\mathrm{g/cm^3}$ + salt + thermal expansion & measured (see Supplemenatry Sec.~2.1) \\
Water permittivity & $\epsilon_r(T)$ & from Ref.~\citenum{fernandez1997formulation} \\
Height of cell & $250~\mathrm{\mu m}$ & known \\
\hline
\multicolumn{3}{l}{\textit{DNA-coated colloid}} \\
\hline
Colloid radius & $a = 2.5~\mathrm{\mu m}$ &  manufacturer \\
Colloid density & $1.055~\mathrm{g/cm^3}$ + thermal expansion & manufacturer + Ref.~\citenum{zakin1966low} for thermal expansion coefficient \\
Brush density & $\sigma_p = (6.9~\mathrm{nm})^{-2} - (2.7~\mathrm{nm})^{-2}$ & measured separately (see Supplementary \S 2.2) \\
PEO molecular weight & $M_{w} = 6.5 - 34~\mathrm{kg/mol}$ & manufacturer \\
PEO persistence length & $0.368~\mathrm{nm}$ & Refs.~\citenum{oelmeier2012molecular,mark1965configuration}\\
PEO brush length & $15 - 35~\mathrm{nm}$ & measured separately (see Supplementary \S 2.1) \\
DNA strands &  refer to Table~\ref{tab:DNAsquence}   & known \\
DNA persistence length & $1.49~\mathrm{nm}$ & At $140~\mathrm{mM}$ salt concentration~\cite{chen2012ionic}  \\
\hline
\multicolumn{3}{l}{\textit{DNA-coated glass surface}} \\
\hline
Brush density & $\sigma_g = (9.4~\mathrm{nm}^2)^{-1}$ & fitted on 1 system and kept constant \\
DNA strands &  refer to Table~\ref{tab:DNAsquence} & known \\
\hline
\multicolumn{3}{l}{\textit{Apparatus}} \\
\hline
Shot noise photon count & $1000$ & known \\
\end{tabular}
\caption{\label{tab:parameters} System parameters used in modeling calculations and method of acquisition}
\end{table}

\clearpage

\end{document}